\documentclass[copyright,creativecommons,sharealike]{eptcs}
\usepackage[utf8]{inputenc}

\title{Recursive Definitions of Monadic Functions}
\newcommand{\titlerunning}{Recursive Definitions of Monadic Functions}

\author{Alexander Krauss
\institute{Technische Universität München, Institut für Informatik}
\email{\url{http://www.in.tum.de/~krauss}}
}
\newcommand{\authorrunning}{A. Krauss}

\DeclareSymbolFont{letters}{OML}{txmi}{m}{it} 

\usepackage[english]{babel}

\usepackage{amsthm}
\usepackage{amsmath}
\theoremstyle{remark}
\newtheorem{example}{Example}

\usepackage{mathpartir}

\usepackage{color}
\definecolor{linkcolor}{rgb}{0,0,0.5}
\hypersetup{colorlinks=true,linkcolor=linkcolor,citecolor=linkcolor,filecolor=linkcolor,pagecolor=linkcolor,urlcolor=linkcolor}

\usepackage{Mrec_Paper/generated/isabelle,Mrec_Paper/generated/isabellesym}
\isabellestyle{it}
\isadroptag{theory}
\renewcommand{\isastyle}{\isastyleminor}

\newcommand{\isasymisub}{\ensuremath{<\!\mid}}
\newcommand{\isasymislub}{\ensuremath{\ll\!\mid}}

{\begin{quote}\begin{tabular}{@{}l@{}c@{}l}}%
{\end{tabular}\end{quote}}
\newenvironment{eqns}%
{\begin{tabular}[t]{@{}l@{}c@{}l}}%
{\end{tabular}}

\newcommand{\blockrule}[3]{%
\inferrule{\mbox{\begin{tabular}{p{#1}}#2\end{tabular}}}{\mbox{#3}}}

\begin{document}

\maketitle

\begin{abstract}
  
Using standard domain-theoretic fixed-points, we present an approach
for defining recursive functions that are formulated in monadic style.
The method works both in the simple option monad and the
state-exception monad of Isabelle/HOL's imperative programming
extension, which results in a convenient definition principle for
imperative programs, which were previously hard to define.

For such monadic functions, the recursion equation can always be
derived without preconditions, even if the function is partial. The
construction is easy to automate, and convenient induction principles
can be derived automatically.

\end{abstract}

\begin{isabellebody}%
\def\isabellecontext{Intro}%
\isadelimtheory
\endisadelimtheory
\isatagtheory
\isacommand{theory}\isamarkupfalse%
\ Intro\isanewline
\isakeyword{imports}\ Imperative{\isacharunderscore}Heap\ Linked{\isacharunderscore}Lists\ Setup\isanewline
\isakeyword{begin}%
\endisatagtheory
{\isafoldtheory}%
\isadelimtheory
\endisadelimtheory
\isamarkupsection{Introduction%
}
\isamarkuptrue%
\begin{isamarkuptext}%
Tool support for non-primitive recursion in interactive
theorem provers has made good progress in the last years. Although the
base logic of most proof assistants has no support for general
recursion or partial functions, tools exist to reduce such definitions
to more basic principles in an automated and mostly transparent way
\cite{wfrec,coqrecdef,Krauss09_funs,Bertot2008}.

This paper discusses a class of definitions which are not
yet supported by any existing tool: imperative computations wrapped up
  in a state monad.  These functions present a challenge, since the
actual structure of the recursion is not directly visible from the
definition itself, but depends on the implicit state argument.
Before considering imperative programs, we first develop our approach
in the simpler option monad. The method can then be extended to state monads
without much difficulty.%
\end{isamarkuptext}%
\isamarkuptrue%
\isamarkupsubsection{Notation%
}
\isamarkuptrue%
\begin{isamarkuptext}%
We work in the
  setting of Isabelle/HOL \cite{tutorial}, which implements a variant
  of classical higher-order logic extended with type classes and
  overloading. Its syntax mostly conforms to standard mathematical
  notation, except for a few idiosyncracies that arise from the generic nature of the
  Isabelle system, notably the two versions of implication (\isa{{\isasymlongrightarrow}} and \isa{{\isasymLongrightarrow}}) and universal quantification (\isa{{\isasymforall}} and
  \isa{{\isasymAnd}}), corresponding to the meta and the object
  level. The reader can safely treat them as interchangeable for the
  purpose of this paper. Function types are written using \isa{{\isasymRightarrow}}, and
  other basic types include \isa{bool}, \isa{nat} and \isa{\isasymalpha\ list}.%
\end{isamarkuptext}%
\isamarkuptrue%
\isamarkupsubsection{Function Definitions in Isabelle/HOL%
}
\isamarkuptrue%
\begin{isamarkuptext}%
To explain the problem, we briefly outline the state of the
  art concerning function definitions in Isabelle/HOL \cite{tutorial}. 
  The definition
  facilities (commonly called the \emph{function package} \cite{Krauss09_funs,krauss_phd})
  work as follows:

  From the specification of a partial function given by a recursive
  equation \isa{f\ x\ {\isacharequal}\ F\ f\ x}, the function package produces a
  total function \isa{f\ {\isacharcolon}{\isacharcolon}\ \isasymsigma\ {\isasymRightarrow}\ \isasymtau}, together with a domain
  predicate \isa{dom\ {\isacharcolon}{\isacharcolon}\ \isasymsigma\ {\isasymRightarrow}\ bool}, which models the set of values
  \isa{x} where the recursion
  terminates. For values outside the domain, we can still write
  the term \isa{f\ x}, but its value may be unknown or meaningless.
  The original recursive specification is then derived as a theorem,
  constrained by the domain condition:
  \begin{quote}\isa{dom\ x\ {\isasymLongrightarrow}\ f\ x\ {\isacharequal}\ F\ f\ x}\end{quote} 
  This condition can be removed by proving
  that the function is total, i.e., \isa{{\isasymforall}x{\isachardot}\ dom\ x}.  An
  induction rule is also derived, which is guarded by the domain
  condition as well.

  The use of a predicate to describe terminating inputs is shared by
  various related
  approaches \cite{ConstableM85,dubois98,bovecapretta05,Greve09_defminterm}.
  In our classical simply-typed setting, it has the advantage
  that the function can be used syntactically as if it were total.  In
  particular, one can treat a total function as partial
  temporarily, until its termination proof is finished and the
  predicate can be discarded. This has proved very useful for nested
  recursion.  However, for truly partial functions the domain
  condition does not go away and has to be dealt with in proofs.%
\end{isamarkuptext}%
\isamarkuptrue%
\isamarkupsubsection{Imperative Functional Programs in Isabelle/HOL%
}
\isamarkuptrue%
\label{sec:imphol}
\begin{isamarkuptext}%
\emph{Imperative HOL} \cite{imperativeFP} is an extension of
  Isabelle/HOL
  which allows modeling and reasoning about programs that manipulate a
  heap. It defines a type \isa{heap}, which models a store where
  references can be allocated and updated.
  (The details behind the type \isa{heap} are omitted and can be safely ignored.)
  \begin{quote}
    \isa{{\isachardoublequote}new{\isacharunderscore}ref\ {\isacharcolon}{\isacharcolon}\ heap\ {\isasymRightarrow}\ \isasymalpha\ ref\ {\isasymtimes}\ heap{\isachardoublequote}}\\
    \isa{{\isachardoublequote}get{\isacharunderscore}ref\ {\isacharcolon}{\isacharcolon}\ \isasymalpha\ ref\ {\isasymRightarrow}\ heap\ {\isasymRightarrow}\ \isasymalpha{\isachardoublequote}}\\
    \isa{{\isachardoublequote}set{\isacharunderscore}ref\ {\isacharcolon}{\isacharcolon}\ \isasymalpha\ ref\ {\isasymRightarrow}\ \isasymalpha\ {\isasymRightarrow}\ heap\ {\isasymRightarrow}\ heap{\isachardoublequote}}
  \end{quote}
  Heap-modifying programs are modelled as monadic computations in
  the so-called \emph{heap monad}:
\begin{quote}
  \isa{\isacommand{datatype}\ \isasymalpha\ Heap\ {\isacharequal}\ Heap\ {\isacharparenleft}heap\ {\isasymRightarrow}\ {\isacharparenleft}\isasymalpha\ {\isacharplus}\ exception{\isacharparenright}\ {\isasymtimes}\ heap{\isacharparenright}}\\[4pt]
  \isa{{\isachardoublequote}return\ {\isacharcolon}{\isacharcolon}\ \isasymalpha\ {\isasymRightarrow}\ \isasymalpha\ Heap{\isachardoublequote}}\\
  \isa{return\ x\ {\isacharequal}\ Heap\ {\isacharparenleft}Pair\ {\isacharparenleft}Inl\ x{\isacharparenright}{\isacharparenright}}\\[4pt]
  \isa{{\isasymguillemotright}{\isacharequal}\ {\isacharcolon}{\isacharcolon}\ \isasymalpha\ Heap\ {\isasymRightarrow}\ {\isacharparenleft}\isasymalpha\ {\isasymRightarrow}\ \isasymbeta\ Heap{\isacharparenright}\ {\isasymRightarrow}\ \isasymbeta\ Heap}\\
  \mbox{\isa{f\ {\isasymguillemotright}{\isacharequal}\ g\ {\isacharequal}\ Heap\ {\isacharparenleft}{\isasymlambda}h{\isachardot}\ case\ exec\ f\ h\ of\ {\isacharparenleft}Inl\ x{\isacharcomma}\ h{\isacharprime}{\isacharparenright}\ {\isasymRightarrow}\ exec\ {\isacharparenleft}g\ x{\isacharparenright}\ h{\isacharprime}\ {\isacharbar}\ {\isacharparenleft}Inr\ exn{\isacharcomma}\ h{\isacharprime}{\isacharparenright}\ {\isasymRightarrow}\ {\isacharparenleft}Inr\ exn{\isacharcomma}\ h{\isacharprime}{\isacharparenright}{\isacharparenright}}}
\end{quote}
  where \isa{exec\ {\isacharparenleft}Heap\ f{\isacharparenright}\ {\isacharequal}\ f}.
  This is nothing more than a state-exception monad, whose state type
  is \isa{heap}. The singleton type \isa{exception} is a simplistic way of
  modelling irrecoverable failure of the computation. The primitive heap operations
  are straightforwardly lifted to monadic operations with the
  following types:
  \begin{quote}
    \isa{{\isachardoublequote}Ref{\isachardot}new\ {\isacharcolon}{\isacharcolon}\ \isasymalpha\ {\isasymRightarrow}\ \isasymalpha\ ref\ Heap{\isachardoublequote}}\\
    \isa{{\isachardoublequote}Ref{\isachardot}lookup\ {\isacharcolon}{\isacharcolon}\ \isasymalpha\ ref\ {\isasymRightarrow}\ \isasymalpha\ Heap{\isachardoublequote}}\\
    \isa{{\isachardoublequote}Ref{\isachardot}update\ {\isacharcolon}{\isacharcolon}\ \isasymalpha\ ref\ {\isasymRightarrow}\ \isasymalpha\ {\isasymRightarrow}\ unit\ Heap{\isachardoublequote}}
  \end{quote}
  
  We abbreviate \isa{{\isachardoublequote}Ref{\isachardot}lookup\ r{\isachardoublequote}} by \isa{{\isacharbang}r} and
  \isa{{\isachardoublequote}Ref{\isachardot}update\ r\ x{\isachardoublequote}} by \isa{r\ {\isacharcolon}{\isacharequal}\ x}. Moreover, we use a do-notation similar to
  Haskell, i.e., \isa{do\ x\ {\isasymleftarrow}\ f{\isacharsemicolon}\ g\ x\ done} abbreviates
  \isa{f\ {\isasymguillemotright}{\isacharequal}\ {\isacharparenleft}{\isasymlambda}x{\isachardot}\ g\ x{\isacharparenright}}.
  For example, here is a data type of heap-allocated linked lists, and
  a function \isa{{\isachardoublequote}traverse\ {\isacharcolon}{\isacharcolon}\ \isasymalpha\ node\ {\isasymRightarrow}\ \isasymalpha\ list\ Heap{\isachardoublequote}} that traverses a linked list and turns it into an
  ordinary list:
\begin{quote}\parskip=0pt
  \isa{\isacommand{datatype}\ \isasymalpha\ node\ {\isacharequal}\ Empty\ {\isacharbar}\ Node\ \isasymalpha\ {\isacharparenleft}\isasymalpha\ node\ ref{\isacharparenright}}
\end{quote}
\begin{quote}\parskip=0pt
  \isa{traverse\ Empty\ {\isacharequal}\ return\ {\isacharbrackleft}{\isacharbrackright}\isasep\isanewline%
traverse\ {\isacharparenleft}Node\ x\ r{\isacharparenright}\ {\isacharequal}\ do\ tl\ {\isasymleftarrow}\ {\isacharbang}r{\isacharsemicolon}\isanewline
\isaindent{traverse\ {\isacharparenleft}Node\ x\ r{\isacharparenright}\ {\isacharequal}\ do\ }xs\ {\isasymleftarrow}\ traverse\ tl{\isacharsemicolon}\isanewline
\isaindent{traverse\ {\isacharparenleft}Node\ x\ r{\isacharparenright}\ {\isacharequal}\ do\ }return\ {\isacharparenleft}x\ {\isacharhash}\ xs{\isacharparenright}\isanewline
\isaindent{traverse\ {\isacharparenleft}Node\ x\ r{\isacharparenright}\ {\isacharequal}\ }done}
\end{quote}
The semantics of a monadic computation \isa{t} is given by a
relation \isa{{\isasymlbrakk}t{\isasymrbrakk}}, where \isa{{\isacharparenleft}h{\isacharcomma}\ h{\isacharprime}{\isacharcomma}\ y{\isacharparenright}\ {\isasymin}\ {\isasymlbrakk}\ t\ {\isasymrbrakk}} expresses that
if the computation is executed on heap \isa{h}, then no
exception occurs and \isa{y} and \isa{h{\isacharprime}} are the result
value and the new heap, respectively.

By a slight extension of Isabelle's code
generator \cite{HaftmannN2010}, the monadic terms can be translated to ML (using
imperative features) or Haskell (using the ST monad).%
\end{isamarkuptext}%
\isamarkuptrue%
\isamarkupsubsection{The Catch: Recursive Monadic Definitions%
}
\isamarkuptrue%
\begin{isamarkuptext}%
But there is a catch! While much of Isabelle's reasoning infrastructure can
  be used for monadic programs as well, the function package
  cannot cope with functions as simple as \isa{traverse} above.

  There are several aspects of the problem. First, the function
  package looks at the arguments of the function to construct the
  domain predicate. However, the termination of the function also
  depends on the heap, which is hidden behind the state monad and not
  a direct argument of the function.

  While this could be solved by breaking the monad abstraction and
  making the heap a normal argument to the function (which would
  result in very messy code), 
  the second problem is that \isa{traverse} is inherently partial,
  as the pointer structure on the heap may be cyclic. 
  Thus, the function package could only produce conditional equations,
  and one would lose the possibility to use the code generator, which
  only works for unconditional equations.  This limitation of code
  generation also applies to other partial functions, but in
  Imperative HOL, where partiality is ubiquitous, it is especially
  problematic.

  The paper on Imperative HOL \cite{imperativeFP} already observes these shortcomings and
  provides a workaround using a recursion combinator \isa{MREC} that can express a common
  case of definitions, including \isa{traverse}. However, the lack
  of tool support soon becomes a show-stopper as soon as more complex
  function definitions are needed.
  
  For example, in ongoing work, Bulwahn is formalizing an imperative
  version of unification. Some of his functions do not fall under the
  scheme of \isa{MREC}, and had to be defined manually in a tedious
  and error-prone process.  This paper aims to simplify this task by
  providing a simpler approach and some automation.%
\end{isamarkuptext}%
\isamarkuptrue%
\isamarkupsubsection{A Solution using Domain Theory%
}
\isamarkuptrue%
\begin{isamarkuptext}%
The approach that we take is to abandon well-founded
  recursion for this task, and resort to domain theory instead, which
  can express very general recursions over complete partial orders.

  The central trick is that the heap monad can be turned into a
  pointed complete partial order (pcpo) by using the exceptions as a
  bottom value. Then, any monadic expression built up from pure terms
  and primitive operations, composed with return and bind operation is
  continuous by construction. This means that the standard least
  fixed-point construction can be used to obtain the function, and
  that the recursive equation can be derived without preconditions.

  A well-developed formalization of domain-theoretic concepts is
  available with Isabelle/HOLCF \cite{holcf}. While we
  build on these concepts as a foundation, the constructions are not
  exposed to the user.%
\end{isamarkuptext}%
\isamarkuptrue%
\isamarkupsubsection{Related Work%
}
\isamarkuptrue%
\begin{isamarkuptext}%
The most closely related work is by Bertot and Komendantsky
\cite{Bertot2008}, who use fixed-points in flat pcpos to define
partial recursive functions in Coq \cite{coqart}, augmented with some
classical axioms. They also show that their extension preserves the
possibility to extract programs from Coq developments and provide a
Coq command that automates the definition process.

This work builds on the same foundations, but goes beyond it in
two points:
\begin{enumerate}
\item We manage to provide an induction principle, which enables
reasoning about the function without having to rely on the underlying
iterative construction.  This permits induction proofs at a higher
level of abstraction, which leads to simpler proofs.

\item By generalizing the approach slightly, we do not only deal with
  the flat domain of the option type, but can also handle
  other situations such as the heap monad. This elegantly
  solves the problem of making recursive definitions of programs in
  Imperative HOL.
\end{enumerate}

Using the monad abstraction to encapsulate partiality is also not new:
In the context of constructive type theory, Capretta
\cite{Capretta2005} and Megacz \cite{Megacz2007} describe monads that
model non-terminating computations coinductively.  In a classical
logic like Isabelle/HOL, however, the much simpler option monad, which
simply adds an extra element to express undefinedness, is sufficient.


\paragraph{The rest of this paper is structured as follows.}
We first introduce the basic preliminaries from Isabelle/HOLCF (Sect.~\ref{holcf}).
Then we show how recursive function definitions can be
automated in the option monad, which is the simplest setting for our
approach (Sect.\ref{recursion}). Sect.~\ref{induction} discusses the automated generation of
induction rules from the general fixed-point induction principle.
Then we move back to the imperative heap monad (Sect.~\ref{heap_recursion}), which is the original
motivation for this work. It will be seen that the technique
generalizes easily to that more interesting case, and we present
a more realistic example.
We compare the method to the domain-predicate-based approach and discuss limitations
and other issues in Sect.~\ref{discussion}.%
\end{isamarkuptext}%
\isamarkuptrue%
\isadelimtheory
\endisadelimtheory
\isatagtheory
\isacommand{end}\isamarkupfalse%
\endisatagtheory
{\isafoldtheory}%
\isadelimtheory
\endisadelimtheory
\end{isabellebody}%

%
\begin{isabellebody}%
\def\isabellecontext{HOLCF{\isacharunderscore}Intro}%
\isadelimtheory
\endisadelimtheory
\isatagtheory
\isacommand{theory}\isamarkupfalse%
\ HOLCF{\isacharunderscore}Intro\isanewline
\isakeyword{imports}\ Imperative{\isacharunderscore}Heap\isanewline
\isakeyword{begin}%
\endisatagtheory
{\isafoldtheory}%
\isadelimtheory
\endisadelimtheory
\isamarkupsection{Isabelle/HOL and Isabelle/HOLCF%
}
\isamarkuptrue%
\label{holcf}
\isadelimproof
\endisadelimproof
\isatagproof
\endisatagproof
{\isafoldproof}%
\isadelimproof
\endisadelimproof
\begin{isamarkuptext}%
Isabelle/HOLCF is a definitional extension of Isabelle/HOL with
  domain-theoretical concepts from the LCF system \cite{Gordon79LCF}. Originally
  developed by Regensburger \cite{Regensburger95,Regensburger_phd},
  its design was later improved by M\"uller et al. \cite{holcf},
  notably by the consequent use of type classes. In recent years, the
  library has been maintained and extended by Huffman
  \cite{Huffman2008,Huffman2009}.

  HOLCF defines a type class of complete partial orders (cpos) \isa{{\isasymsqsubseteq}}, based on which the standard notions of chain, (least) upper
  bounds, and continuity are defined as in Fig.~\ref{fig:defs}.
  Note that the expression \isa{{\isasymSqunion}\ i{\isachardot}\ Y\ i} only denotes a meaningful
  value if a least upper bound actually exists. Otherwise its value is
  arbitrary. A cpo is \emph{pointed} if it has a least element,
  written \isa{{\isasymbottom}}. Pointed cpos are also called pcpos.

\begin{figure}
\begin{quote}
  \isa{{\isachardoublequote}chain\ {\isacharcolon}{\isacharcolon}\ {\isacharparenleft}nat\ {\isasymRightarrow}\ \isasymalpha{\isacharparenright}\ {\isasymRightarrow}\ bool{\isachardoublequote}}\\
  \isa{chain\ Y\ {\isasymlongleftrightarrow}\ {\isacharparenleft}{\isasymforall}i{\isachardot}\ Y\ i\ {\isasymsqsubseteq}\ Y\ {\isacharparenleft}Suc\ i{\isacharparenright}{\isacharparenright}}\\[4pt]
  \isa{{\isachardoublequote}range\ {\isacharcolon}{\isacharcolon}\ {\isacharparenleft}\isasymalpha\ {\isasymRightarrow}\ \isasymbeta{\isacharparenright}\ {\isasymRightarrow}\ \isasymbeta\ set{\isachardoublequote}}\\
  \isa{range\ f\ {\isacharequal}\ {\isacharbraceleft}y{\isachardot}\ {\isasymexists}x{\isachardot}\ y\ {\isacharequal}\ f\ x{\isacharbraceright}}\\[4pt]
  \isa{{\isasymisub}\ {\isacharcolon}{\isacharcolon}\ \isasymalpha\ set\ {\isasymRightarrow}\ \isasymalpha\ {\isasymRightarrow}\ bool}\\
  \isa{S\ {\isasymisub}\ x\ {\isasymlongleftrightarrow}\ {\isacharparenleft}{\isasymforall}y{\isachardot}\ y\ {\isasymin}\ S\ {\isasymlongrightarrow}\ y\ {\isasymsqsubseteq}\ x{\isacharparenright}}\\[4pt]
  \isa{{\isasymislub}\ {\isacharcolon}{\isacharcolon}\ \isasymalpha\ set\ {\isasymRightarrow}\ \isasymalpha\ {\isasymRightarrow}\ bool}\\
  \isa{S\ {\isasymislub}\ x\ {\isasymlongleftrightarrow}\ S\ {\isasymisub}\ x\ {\isasymand}\ {\isacharparenleft}{\isasymforall}u{\isachardot}\ S\ {\isasymisub}\ u\ {\isasymlongrightarrow}\ x\ {\isasymsqsubseteq}\ u{\isacharparenright}}\\[4pt]
  \isa{{\isasymSqunion}\ {\isacharcolon}{\isacharcolon}\ \isasymalpha\ set\ {\isasymRightarrow}\ \isasymalpha}\\
  \isa{{\isacharparenleft}{\isasymSqunion}\ i{\isachardot}\ Y\ i{\isacharparenright}\ {\isacharequal}\ {\isacharparenleft}THE\ x{\isachardot}\ range\ Y\ {\isasymislub}\ x{\isacharparenright}}\\[4pt]
  \isa{{\isachardoublequote}cont\ {\isacharcolon}{\isacharcolon}\ {\isacharparenleft}\isasymalpha\ {\isasymRightarrow}\ \isasymbeta{\isacharparenright}\ {\isasymRightarrow}\ bool{\isachardoublequote}}\\
  \isa{cont\ f\ {\isasymlongleftrightarrow}\ {\isacharparenleft}{\isasymforall}Y{\isachardot}\ chain\ Y\ {\isasymlongrightarrow}\ range\ {\isacharparenleft}{\isasymlambda}i{\isachardot}\ f\ {\isacharparenleft}Y\ i{\isacharparenright}{\isacharparenright}\ {\isasymislub}\ f\ {\isacharparenleft}{\isasymSqunion}\ i{\isachardot}\ Y\ i{\isacharparenright}{\isacharparenright}}
\end{quote}
\caption{Basic definitions of HOLCF}\label{fig:defs}
\end{figure}

  One of the basic results of domain theory is a fixed-point theorem,
  proving that continuous functions always have a fixed-point that can
  be reached by iteration:
\begin{equation}
  \isa{cont\ F\ {\isasymLongrightarrow}\ fixp\ F\ {\isacharequal}\ F\ {\isacharparenleft}fixp\ F{\isacharparenright}}\tag{\sc Fixp}\label{fixp_eq}
\end{equation}
Here, \isa{fixp\ F\ {\isacharequal}\ {\isacharparenleft}{\isasymSqunion}\ i{\isachardot}\ F\isactrlbsup i\isactrlesup \ {\isasymbottom}{\isacharparenright}}, and \isa{F\isactrlbsup i\isactrlesup } denotes iterated function
application.

As one of its main features, HOLCF then introduces a type of
continuous functions, written \isa{\isasymalpha\ {\isasymrightarrow}\ \isasymbeta}. While this helps to
automate many continuity proofs by turning them into type checking, it
also destroys the compatibility with regular Isabelle/HOL
developments.  Since we are trying to simplify function definitions in
HOL, we do not use the continuous function space.

More generally, while this work uses concepts from HOLCF, it is
important to note that this is completely transparent and just part
of the internal construction. A user of our tool does not have to know
  domain theory or its formalization in HOLCF.%
\end{isamarkuptext}%
\isamarkuptrue%
\isadelimtheory
\endisadelimtheory
\isatagtheory
\isacommand{end}\isamarkupfalse%
\endisatagtheory
{\isafoldtheory}%
\isadelimtheory
\endisadelimtheory
\end{isabellebody}%

%
\begin{isabellebody}%
\def\isabellecontext{Recursion}%
\isadelimtheory
\endisadelimtheory
\isatagtheory
\isacommand{theory}\isamarkupfalse%
\ Recursion\isanewline
\isakeyword{imports}\ Setup\ Option{\isacharunderscore}Examples\isanewline
\isakeyword{begin}%
\endisatagtheory
{\isafoldtheory}%
\isadelimtheory
\endisadelimtheory
\isamarkupsection{Recursion in the Option Monad%
}
\isamarkuptrue%
\label{recursion}
\begin{isamarkuptext}%
This section shows how to define partial functions in the option
  monad. It mainly recalls the standard fixed-point construction also
  used by Bertot and Komendantsky \cite{Bertot2008}, and shows how it
  is automated in Isabelle/HOL. Later we will generalize it to the
  heap monad.

  We start from the standard option type in Isabelle/HOL, together with
  the monad operations:
  \begin{quote}
  \isa{\isacommand{datatype}\ \isasymalpha\ option\ {\isacharequal}\ None\ {\isacharbar}\ Some\ \isasymalpha}\\[1ex]
  \isa{return\ x\ {\isacharequal}\ Some\ x}\\[1ex]
  \begin{eqns}
  \isa{None\ {\isasymguillemotright}{\isacharequal}\ f} & \isa{\ {\isacharequal}\ } & \isa{None}\\
  \isa{Some\ y\ {\isasymguillemotright}{\isacharequal}\ f} & \isa{\ {\isacharequal}\ } & \isa{f\ y}
  \end{eqns}
  \end{quote}
  This monad is known to Haskell programmers as the Maybe monad and models
  computations with failure. However, it can also be
  regarded as a (flat) pcpo, where \isa{{\isasymbottom}\ {\isacharequal}\ None}.

  This basically (ab)uses \isa{None} as the result of a
  non-terminating computation. The fixed-point law can thus be used to solve 
  recursive equations \isa{f\ x\ {\isacharequal}\ F\ f\ x}, provided that the functional involved is continuous:
  \begin{enumerate}\parskip=0pt
  \item Prove that the functional \isa{F} is continuous.
  \item Define \isa{f\ {\isacharequal}\ fixp\ F}.
  \item Conclude the equation \isa{f\ x\ {\isacharequal}\ F\ f\ x} using the fixed-point theorem \eqref{fixp_eq}.
  \end{enumerate}

  Now the primary observation is that if the function is
  written in monadic style, continuity holds by construction and can be proved 
  automatically following the term structure using the rules given in Fig.~\ref{fig:cont}.
  \begin{figure}
  \begin{quote}
  \begin{tabular}{lcr}
  \isa{{\isacharparenleft}{\isasymAnd}y{\isachardot}\ cont\ {\isacharparenleft}{\isasymlambda}x{\isachardot}\ f\ x\ y{\isacharparenright}{\isacharparenright}\ {\isasymLongrightarrow}\ cont\ f}&&{\sc (Lam)}\\
  \isa{cont\ f\ {\isasymLongrightarrow}\ {\isacharparenleft}{\isasymAnd}y{\isachardot}\ cont\ {\isacharparenleft}g\ y{\isacharparenright}{\isacharparenright}\ {\isasymLongrightarrow}\ cont\ {\isacharparenleft}{\isasymlambda}x{\isachardot}\ do\ y\ {\isasymleftarrow}\ f\ x{\isacharsemicolon}\ g\ y\ x\ done{\isacharparenright}}&&{\sc (Bind)}\\
  \isa{cont\ {\isacharparenleft}{\isasymlambda}x{\isachardot}\ c{\isacharparenright}}&& {\sc (Const)}\\
  \isa{cont\ {\isacharparenleft}{\isasymlambda}f{\isachardot}\ f\ x{\isacharparenright}}&& {\sc (Rec)}\\
  \isa{cont\ f\ {\isasymLongrightarrow}\ cont\ g\ {\isasymLongrightarrow}\ cont\ {\isacharparenleft}{\isasymlambda}x{\isachardot}\ if\ b\ then\ f\ x\ else\ g\ x{\isacharparenright}}&& {\sc (If)}
  \end{tabular}
  \end{quote}
  \caption{Continuity rules}\label{fig:cont}
  \end{figure}

  \begin{example}As an artificial example,
  assume some fixed function \isa{{\isachardoublequote}step\ {\isacharcolon}{\isacharcolon}\ nat\ {\isasymRightarrow}\ nat{\isachardoublequote}}
  and assume that we want to define a function 
  \isa{trace\ {\isacharcolon}{\isacharcolon}\ nat\ {\isasymRightarrow}\ nat\ list} that iterates the step
  function, until it returns zero, keeping all even values in a list.
  Here is how the function could be written in ML.\newpage

  \begin{quote}\upshape\tt
  fun trace n = \\
  \mbox{}~~if n = 0 then []\\
  \mbox{}~~else if even n then n ::~trace (step n) else trace (step n)
  \end{quote}\end{example}

  \noindent
  The function is partial and asserting this equation directly in Isabelle/HOL would
  be unsound. However, we can define its monadic counterpart
  with return type \isa{nat\ list\ option}: (Note that \isa{{\isacharhash}} is
  Isabelle's way of spelling the constructor for non-empty lists)
  \begin{quote}\parskip=0pt
  \isa{trace\ n\ {\isacharequal}\isanewline
{\isacharparenleft}if\ n\ {\isacharequal}\ {\isadigit{0}}\ then\ return\ {\isacharbrackleft}{\isacharbrackright}\isanewline
\isaindent{{\isacharparenleft}}else\ do\ tl\ {\isasymleftarrow}\ trace\ {\isacharparenleft}step\ n{\isacharparenright}{\isacharsemicolon}\isanewline
\isaindent{{\isacharparenleft}else\ do\ }{\isacharparenleft}if\ even\ n\ then\ return\ {\isacharparenleft}n\ {\isacharhash}\ tl{\isacharparenright}\ else\ return\ tl{\isacharparenright}\isanewline
\isaindent{{\isacharparenleft}else\ }done{\isacharparenright}}
  \end{quote}%
\end{isamarkuptext}%
\isamarkuptrue%
\isadelimproof
\endisadelimproof
\isatagproof
\begin{isamarkuptxt}%
The following step-by-step proof shows that continuity
of the functional is easily proved in a completely syntax-directed way.
The proof obligation is as follows.
  \begin{quote}\parskip=0pt\isa{\ {\isadigit{1}}{\isachardot}\ cont\ {\isacharparenleft}{\isasymlambda}trace\ n{\isachardot}\isanewline
\isaindent{\ {\isadigit{1}}{\isachardot}\ cont\ {\isacharparenleft}\ \ \ }if\ n\ {\isacharequal}\ {\isadigit{0}}\ then\ return\ {\isacharbrackleft}{\isacharbrackright}\isanewline
\isaindent{\ {\isadigit{1}}{\isachardot}\ cont\ {\isacharparenleft}\ \ \ }else\ do\ tl\ {\isasymleftarrow}\ trace\ {\isacharparenleft}step\ n{\isacharparenright}{\isacharsemicolon}\isanewline
\isaindent{\ {\isadigit{1}}{\isachardot}\ cont\ {\isacharparenleft}\ \ \ else\ do\ }{\isacharparenleft}if\ even\ n\ then\ return\ {\isacharparenleft}n\ {\isacharhash}\ tl{\isacharparenright}\ else\ return\ tl{\isacharparenright}\isanewline
\isaindent{\ {\isadigit{1}}{\isachardot}\ cont\ {\isacharparenleft}\ \ \ else\ }done{\isacharparenright}}\end{quote}
  We first move the lambda bound argument out using rule {\sc (Lam)}:%
\end{isamarkuptxt}%
\isamarkuptrue%
\begin{isamarkuptxt}%
\begin{quote}\parskip=0pt\isa{\ {\isadigit{1}}{\isachardot}\ {\isasymAnd}n{\isachardot}\ cont\ {\isacharparenleft}{\isasymlambda}trace{\isachardot}\ if\ n\ {\isacharequal}\ {\isadigit{0}}\ then\ return\ {\isacharbrackleft}{\isacharbrackright}\isanewline
\isaindent{\ {\isadigit{1}}{\isachardot}\ {\isasymAnd}n{\isachardot}\ cont\ {\isacharparenleft}{\isasymlambda}trace{\isachardot}\ }else\ do\ tl\ {\isasymleftarrow}\ trace\ {\isacharparenleft}step\ n{\isacharparenright}{\isacharsemicolon}\isanewline
\isaindent{\ {\isadigit{1}}{\isachardot}\ {\isasymAnd}n{\isachardot}\ cont\ {\isacharparenleft}{\isasymlambda}trace{\isachardot}\ else\ do\ }{\isacharparenleft}if\ even\ n\ then\ return\ {\isacharparenleft}n\ {\isacharhash}\ tl{\isacharparenright}\ else\ return\ tl{\isacharparenright}\isanewline
\isaindent{\ {\isadigit{1}}{\isachardot}\ {\isasymAnd}n{\isachardot}\ cont\ {\isacharparenleft}{\isasymlambda}trace{\isachardot}\ else\ }done{\isacharparenright}}\end{quote}
Applying rule {\sc (If)}, we obtain two subgoals:%
\end{isamarkuptxt}%
\isamarkuptrue%
\begin{isamarkuptxt}%
\begin{quote}\parskip=0pt\isa{\ {\isadigit{1}}{\isachardot}\ {\isasymAnd}n{\isachardot}\ cont\ {\isacharparenleft}{\isasymlambda}trace{\isachardot}\ return\ {\isacharbrackleft}{\isacharbrackright}{\isacharparenright}\isanewline
\ {\isadigit{2}}{\isachardot}\ {\isasymAnd}n{\isachardot}\ cont\ {\isacharparenleft}{\isasymlambda}trace{\isachardot}\ do\ tl\ {\isasymleftarrow}\ trace\ {\isacharparenleft}step\ n{\isacharparenright}{\isacharsemicolon}\isanewline
\isaindent{\ {\isadigit{2}}{\isachardot}\ {\isasymAnd}n{\isachardot}\ cont\ {\isacharparenleft}{\isasymlambda}trace{\isachardot}\ do\ }{\isacharparenleft}if\ even\ n\ then\ return\ {\isacharparenleft}n\ {\isacharhash}\ tl{\isacharparenright}\ else\ return\ tl{\isacharparenright}\isanewline
\isaindent{\ {\isadigit{2}}{\isachardot}\ {\isasymAnd}n{\isachardot}\ cont\ {\isacharparenleft}{\isasymlambda}trace{\isachardot}\ }done{\isacharparenright}}\end{quote}
The first goal is trivial as it contains no recursive
  call, and can be discharged with rule {\sc (Const)}. The other goal contains a
  bind, and we decompose it using rule {\sc (Bind)}:%
\end{isamarkuptxt}%
\isamarkuptrue%
\begin{isamarkuptxt}%
\begin{quote}\parskip=0pt\isa{\ {\isadigit{1}}{\isachardot}\ {\isasymAnd}n{\isachardot}\ cont\ {\isacharparenleft}{\isasymlambda}trace{\isachardot}\ trace\ {\isacharparenleft}step\ n{\isacharparenright}{\isacharparenright}\isanewline
\ {\isadigit{2}}{\isachardot}\ {\isasymAnd}n\ tl{\isachardot}\ cont\ {\isacharparenleft}{\isasymlambda}trace{\isachardot}\ if\ even\ n\ then\ return\ {\isacharparenleft}n\ {\isacharhash}\ tl{\isacharparenright}\ else\ return\ tl{\isacharparenright}}\end{quote}
  Now, the first goal is a recursive call, and we apply rule {\sc
  (Rec)}.
  The other goal is again trivially solved using {\sc (Const)}.%
\end{isamarkuptxt}%
\isamarkuptrue%
\endisatagproof
{\isafoldproof}%
\isadelimproof
\endisadelimproof
\isadelimtheory
\endisadelimtheory
\isatagtheory
\endisatagtheory
{\isafoldtheory}%
\isadelimtheory
\endisadelimtheory
\end{isabellebody}%

%
\begin{isabellebody}%
\def\isabellecontext{Induction{\isacharunderscore}Rules}%
\isadelimtheory
\endisadelimtheory
\isatagtheory
\isacommand{theory}\isamarkupfalse%
\ Induction{\isacharunderscore}Rules\isanewline
\isakeyword{imports}\ Recursion\isanewline
\isakeyword{begin}%
\endisatagtheory
{\isafoldtheory}%
\isadelimtheory
\endisadelimtheory
\isamarkupsection{Induction Rules for Partial Correctness%
}
\isamarkuptrue%
\label{induction}
\isadelimproof
\endisadelimproof
\isatagproof
\endisatagproof
{\isafoldproof}%
\isadelimproof
\endisadelimproof
\isadelimproof
\endisadelimproof
\isatagproof
\endisatagproof
{\isafoldproof}%
\isadelimproof
\endisadelimproof
\isadelimproof
\endisadelimproof
\isatagproof
\endisatagproof
{\isafoldproof}%
\isadelimproof
\endisadelimproof
\begin{isamarkuptext}%
Defining the function and deriving the recursive equation is always
  just half of the problem, since one wants to use induction to
  reason about the function. For total functions, the induction principle
  is a consequence of the termination of the function.
  For partial functions, the function package can generate a similar
  rule, using the domain predicate as a guard.
  In this section, we show how to derive an induction rule for partial
  functions constructed as least fixed-points. 

  HOLCF provides a general fixed-point induction rule:
  \begin{quote}
  \isa{adm\ P\ {\isasymLongrightarrow}\ cont\ F\ {\isasymLongrightarrow}\ P\ {\isasymbottom}\ {\isasymLongrightarrow}\ {\isacharparenleft}{\isasymAnd}f{\isachardot}\ P\ f\ {\isasymLongrightarrow}\ P\ {\isacharparenleft}F\ f{\isacharparenright}{\isacharparenright}\ {\isasymLongrightarrow}\ P\ {\isacharparenleft}fixp\ F{\isacharparenright}}
  \end{quote}
  Besides continuity, which is already proved at definition time, the
  property \isa{P} must hold for \isa{{\isasymbottom}} and it must be
  \emph{admissible}, which means that it can be transferred from
  chains to least upper bounds:
  \begin{quote}
  \isa{adm\ P\ {\isacharequal}\ {\isacharparenleft}{\isasymforall}Y{\isachardot}\ chain\ Y\ {\isasymlongrightarrow}\ {\isacharparenleft}{\isasymforall}i{\isachardot}\ P\ {\isacharparenleft}Y\ i{\isacharparenright}{\isacharparenright}\ {\isasymlongrightarrow}\ P\ {\isacharparenleft}{\isasymSqunion}\ i{\isachardot}\ Y\ i{\isacharparenright}{\isacharparenright}}
  \end{quote}
  
  While this general rule can be used to reason about the 
  function, it is somewhat abstract and not very convenient to
  use. In particular, the admissibility condition must always be proved when applying the
  rule. Although HOLCF can automate such proofs in some cases,
  we would prefer to hide these inconvenient parts of domain theory completely.%
\end{isamarkuptext}%
\isamarkuptrue%
\isamarkupsubsection{Restriction to partial correctness%
}
\isamarkuptrue%
\begin{isamarkuptext}%
It turns out that there is an instance of the general rule 
which is easier to work with. 

If we restrict ourselves to partial correctness properties, i.e.,
showing that the result of the function, when it is defined, satisfies
some predicate, then matters become straightforward.  More precisely,
we replace the predicate \isa{P} with the instance \isa{{\isasymlambda}f{\isachardot}\ {\isasymforall}x\ y{\isachardot}\ f\ x\ {\isacharequal}\ Some\ y\ {\isasymlongrightarrow}\ Q\ x\ y}. Then the admissibility condition can be discharged once
and for all, since this instance is always admissible.

Thus, if \isa{f} is the recursive function, and \isa{F} is the
corresponding functional, the following rule can be derived.
\begin{center}\mbox{}%
\inferrule{\isa{{\isasymAnd}f\ x\ y{\isachardot}\ {\isacharparenleft}{\isasymAnd}z\ r{\isachardot}\ f\ z\ {\isacharequal}\ Some\ r\ {\isasymLongrightarrow}\ Q\ z\ r{\isacharparenright}\ {\isasymLongrightarrow}\ F\ f\ x\ {\isacharequal}\ Some\ y\ {\isasymLongrightarrow}\ Q\ x\ y}}%
{\isa{f\ x\ {\isacharequal}\ Some\ y}\ \isa{{\isasymLongrightarrow}}\ \isa{Q\ x\ y}}
\end{center}

Note that the statement of this rule makes no mentioning of the
iterative fixed-point construction. Presenting this rule to the user
hides the details of this construction, which allows reasoning on a
more abstract level.

\begin{example}
  For example, the instance for \emph{trace} is as follows:
  \begin{center}\mbox{}
\blockrule{0.55\textwidth}%
  {\isa{{\isasymAnd}trace\ n\ ys{\isachardot}\isanewline
\isaindent{\ \ \ }{\isacharparenleft}{\isasymAnd}z\ r{\isachardot}\ trace\ z\ {\isacharequal}\ Some\ r\ {\isasymLongrightarrow}\ Q\ z\ r{\isacharparenright}\ {\isasymLongrightarrow}\isanewline
\isaindent{\ \ \ }{\isacharparenleft}if\ n\ {\isacharequal}\ {\isadigit{0}}\ then\ return\ {\isacharbrackleft}{\isacharbrackright}\isanewline
\isaindent{\ \ \ {\isacharparenleft}}else\ do\ tl\ {\isasymleftarrow}\ trace\ {\isacharparenleft}step\ n{\isacharparenright}{\isacharsemicolon}\isanewline
\isaindent{\ \ \ {\isacharparenleft}else\ do\ }{\isacharparenleft}if\ even\ n\ then\ return\ {\isacharparenleft}n\ {\isacharhash}\ tl{\isacharparenright}\ else\ return\ tl{\isacharparenright}\isanewline
\isaindent{\ \ \ {\isacharparenleft}else\ }done{\isacharparenright}\ {\isacharequal}\isanewline
\isaindent{\ \ \ }Some\ ys\ {\isasymLongrightarrow}\isanewline
\isaindent{\ \ \ }Q\ n\ ys}}%
{\isa{trace\ n\ {\isacharequal}\ Some\ ys}\ \isa{{\isasymLongrightarrow}}\ \isa{Q\ n\ ys}}
\end{center}
\end{example}

\noindent%
\end{isamarkuptext}%
\isamarkuptrue%
\isamarkupsubsection{Induction Rule Refinement%
}
\isamarkuptrue%
\begin{isamarkuptext}%
The raw induction rule as shown above can still be improved.  First,
the control flow in the definition gives rise to three cases, one for
\isa{{\isachardoublequote}n\ {\isacharequal}\ {\isadigit{0}}{\isachardoublequote}}, one for
\isa{even\ n}, and one for \isa{{\isasymnot}\ even\ n}. Second, the
  sequencing using 
\isa{{\isasymguillemotright}{\isacharequal}} can be decomposed, since 
the whole expression is only defined when all relevant subexpressions are
defined. Moreover, the induction hypothesis is likely to be only
useful to prove that the recursive call satisfies the property \isa{Q}. The rule that a user would like to see is roughly the following:
  \begin{center}\mbox{} \blockrule{0.6\textwidth}%
{\isa{Q\ {\isadigit{0}}\ {\isacharbrackleft}{\isacharbrackright}}\\
\isa{{\isasymAnd}n\ tl{\isachardot}\ n\ {\isasymnoteq}\ {\isadigit{0}}\ {\isasymLongrightarrow}\ Q\ {\isacharparenleft}step\ n{\isacharparenright}\ tl\ {\isasymLongrightarrow}\ even\ n\ {\isasymLongrightarrow}\ Q\ n\ {\isacharparenleft}n\ {\isacharhash}\ tl{\isacharparenright}}\\
\isa{{\isasymAnd}n\ tl{\isachardot}\ n\ {\isasymnoteq}\ {\isadigit{0}}\ {\isasymLongrightarrow}\ Q\ {\isacharparenleft}step\ n{\isacharparenright}\ tl\ {\isasymLongrightarrow}\ {\isasymnot}\ even\ n\ {\isasymLongrightarrow}\ Q\ n\ tl}}%
{\isa{trace\ n\ {\isacharequal}\ Some\ ys}\ \isa{{\isasymLongrightarrow}}\ \isa{Q\ n\ ys}}
\end{center}

\noindent To arrive at this simpler form, the following steps are necessary:
\begin{enumerate}
\item Decompose the program structure by splitting the function body into
  smaller steps:
  \begin{itemize}
  \item A premise \isa{{\isachardoublequote}{\isacharparenleft}t\ {\isasymguillemotright}{\isacharequal}\ {\isacharparenleft}{\isasymlambda}x{\isachardot}\ f\ x{\isacharparenright}{\isacharparenright}\ {\isacharequal}\ Some\ y{\isachardoublequote}} is replaced by \isa{t\ {\isacharequal}\ Some\ x} and \isa{f\ x\ {\isacharequal}\ Some\ y}
  \item A premise \isa{{\isachardoublequote}Some\ t\ {\isacharequal}\ Some\ y{\isachardoublequote}} (which arises from a return
  statement) is replaced by \isa{t\ {\isacharequal}\ y}.
  \item Conditionals like \isa{{\isacharparenleft}if\ b\ then\ x\ else\ x{\isacharprime}{\isacharparenright}\ {\isacharequal}\ Some\ y} are
  split up into two cases, with premises \isa{b} and \isa{{\isasymnot}\ b}.
  \end{itemize}
\item Use the induction hypothesis to replace premises of the form
  \isa{f\ z\ {\isacharequal}\ Some\ r} by \isa{Q\ z\ r}. When all occurrences of
  \isa{f} are replaced, the general induction hypothesis can be
  discarded.
\item Clean up the context by substituting premises of the form \isa{v\ {\isacharequal}\ t} where \isa{v} is a variable.
\end{enumerate}
\end{isamarkuptext}%
\isamarkuptrue%
\isadelimtheory
\endisadelimtheory
\isatagtheory
\isacommand{end}\isamarkupfalse%
\endisatagtheory
{\isafoldtheory}%
\isadelimtheory
\endisadelimtheory
\end{isabellebody}%

%
\begin{isabellebody}%
\def\isabellecontext{Recursion{\isacharunderscore}Heap}%
\isadelimtheory
\endisadelimtheory
\isatagtheory
\isacommand{theory}\isamarkupfalse%
\ Recursion{\isacharunderscore}Heap\isanewline
\isakeyword{imports}\ Setup\ Heap{\isacharunderscore}Examples\isanewline
\isakeyword{begin}%
\endisatagtheory
{\isafoldtheory}%
\isadelimtheory
\endisadelimtheory
\isadelimproof
\endisadelimproof
\isatagproof
\endisatagproof
{\isafoldproof}%
\isadelimproof
\endisadelimproof
\isamarkupsection{Recursion in the Heap Monad%
}
\isamarkuptrue%
\label{heap_recursion}
\begin{isamarkuptext}%
We now move from the option monad to the more interesting heap
  monad. In fact, not much of the process has to be adapted.

  \begin{description}
  \item[The heap pcpo.] Unlike the option pcpo, the heap
  pcpo is not flat, since its values represent state transformations.
  The order is defined as \isa{Heap\ f\ {\isasymsqsubseteq}\ Heap\ g\ {\isasymlongleftrightarrow}\ {\isacharparenleft}{\isasymforall}h{\isachardot}\ f\ h\ {\isacharequal}\ bot\ {\isasymor}\ f\ h\ {\isacharequal}\ g\ h{\isacharparenright}}, where \mbox{\isa{bot\ {\isacharequal}\ {\isacharparenleft}Inr\ Exn{\isacharcomma}\ h\isactrlsub {\isadigit{0}}{\isacharparenright}}} for some
  arbitrary but fixed heap \isa{h\isactrlsub {\isadigit{0}}}. This implies \isa{{\isasymbottom}\ {\isacharequal}\ Heap\ {\isacharparenleft}{\isasymlambda}h{\isachardot}\ bot{\isacharparenright}}.

  \item[Recursive definitions.]
  After proving continuity of \isa{{\isasymguillemotright}{\isacharequal}}, which is tedious but
  straightforward, the definition process remains the same as for the
  option monad. We automatically prove continuity of the functional by
  applying the rules from Fig.~\ref{fig:cont} in a syntax-directed
  way. After that, the function can be defined as a fixed-point.

  \item[Induction rule generation.]
  In the induction rule, the condition \isa{f\ x\ {\isacharequal}\ Some\ y} is
  replaced with its counterpart for heap-manipulating programs,
  the condition \isa{{\isacharparenleft}h{\isacharcomma}\ h{\isacharprime}{\isacharcomma}\ y{\isacharparenright}\ {\isasymin}\ {\isasymlbrakk}\ f\ x\ {\isasymrbrakk}} (cf.~Sect.~\ref{sec:imphol}).
  The inductive property \isa{Q} now also refers to the heap before and after the computation.
  As in the option case, we must prove that this partial correctness
  property is always admissible:
  \begin{quote}
  \isa{adm\ {\isacharparenleft}{\isasymlambda}f{\isachardot}\ {\isasymforall}x\ h\ h{\isacharprime}\ y{\isachardot}\ {\isacharparenleft}h{\isacharcomma}\ h{\isacharprime}{\isacharcomma}\ y{\isacharparenright}\ {\isasymin}\ {\isasymlbrakk}\ f\ x\ {\isasymrbrakk}\ {\isasymlongrightarrow}\ Q\ x\ h\ h{\isacharprime}\ y{\isacharparenright}}
  \end{quote}
  \item[Induction rule refinement.] In the refinement process, the program structure is
  decomposed as described above. For the primitive heap operations,
  additional refinement steps are added, which replace them by their
  counterparts with explicit heap. For example, the premise \isa{{\isacharparenleft}h{\isacharcomma}\ h{\isacharprime}{\isacharcomma}\ y{\isacharparenright}\ {\isasymin}\ {\isasymlbrakk}\ {\isacharbang}r\ {\isasymrbrakk}} is replaced by \isa{y\ {\isacharequal}\ get{\isacharunderscore}ref\ r\ h} and \isa{h{\isacharprime}\ {\isacharequal}\ h}.
  \end{description}

  \begin{example}
  The refined induction rule for \isa{traverse} has the following 
  form:
  \begin{center}\mbox{} \blockrule{0.8\textwidth}%
{\isa{{\isasymAnd}h{\isacharprime}{\isachardot}\ Q\ Empty\ h{\isacharprime}\ h{\isacharprime}\ {\isacharbrackleft}{\isacharbrackright}}\\
\isa{{\isasymAnd}h\isactrlisub {\isadigit{1}}\ h\isactrlisub {\isadigit{2}}\ x{\isacharprime}\ r\ n{\isachardot}\ Q\ {\isacharparenleft}get{\isacharunderscore}ref\ r\ h\isactrlisub {\isadigit{1}}{\isacharparenright}\ h\isactrlisub {\isadigit{1}}\ h\isactrlisub {\isadigit{2}}\ n\ {\isasymLongrightarrow}\ Q\ {\isacharparenleft}Node\ x{\isacharprime}\ r{\isacharparenright}\ h\isactrlisub {\isadigit{1}}\ h\isactrlisub {\isadigit{2}}\ {\isacharparenleft}x{\isacharprime}\ {\isacharhash}\ n{\isacharparenright}}}%
{\isa{{\isacharparenleft}h{\isacharcomma}\ h{\isacharprime}{\isacharcomma}\ y{\isacharparenright}\ {\isasymin}\ {\isasymlbrakk}\ traverse\ x\ {\isasymrbrakk}}\ \isa{{\isasymLongrightarrow}}\ \isa{Q\ x\ h\ h{\isacharprime}\ y}}
\end{center}
  \end{example}

\begin{example}
  We now discuss a more realistic example that arises in the
  formalization of an imperative unification algorithm
  mentioned previously.
  We refer to Baader and Nipkow \cite[ch.~4.8]{nipkow_rewriting}.
  for a textbook description of imperative unification.
  To avoid expensive allocations, the algorithm keeps track of
  substitutions by directly updating references in the terms
  themselves.

  In the formalization, heap-allocated mutable terms 
  consist of variables, constants and binary applications:
  \begin{quote}
  \isa{\isacommand{datatype}\ \isasymalpha\ rtrm\ {\isacharequal}\ Var\ \isasymalpha\ {\isacharparenleft}\isasymalpha\ rtrm\ ref\ option{\isacharparenright}\ {\isacharbar}\ Const\ \isasymalpha\ {\isacharbar}\ App\ {\isacharparenleft}\isasymalpha\ rtrm\ ref{\isacharparenright}\ {\isacharparenleft}\isasymalpha\ rtrm\ ref{\isacharparenright}}
  \end{quote}
  The type argument \isa{\isasymalpha} is only used for names. Note
  that variables carry an optional reference cell, which is
  used to mark that a variable has already been assigned some other
  term. Applying a substitution for that variable only requires an
  update of the relevant reference, which also affects other
  occurrences of the same variable, since the reference is shared.
  A value of \isa{None} means that the variable is unassigned.

  We will only show a simple
  part of the unification algorithm, namely the function \isa{occurs}, which
  checks if a variable \isa{r\isactrlisub {\isadigit{1}}} appears in some term \isa{r\isactrlisub {\isadigit{2}}}:
\begin{quote}\parskip=0pt
\isa{occurs\ r\isactrlisub {\isadigit{1}}\ r\isactrlisub {\isadigit{2}}\ {\isacharequal}\isanewline
do\ t\ {\isasymleftarrow}\ {\isacharbang}r\isactrlisub {\isadigit{2}}{\isacharsemicolon}\isanewline
\isaindent{do\ }{\isacharparenleft}case\ t\ of\isanewline
\isaindent{do\ {\isacharparenleft}}Var\ n\ {\isasymsigma}\ {\isasymRightarrow}\isanewline
\isaindent{do\ {\isacharparenleft}\ \ }if\ r\isactrlisub {\isadigit{1}}\ {\isacharequal}\ r\isactrlisub {\isadigit{2}}\ then\ return\ True\isanewline
\isaindent{do\ {\isacharparenleft}\ \ }else\ case\ {\isasymsigma}\ of\ None\ {\isasymRightarrow}\ return\ False\ {\isacharbar}\ Some\ r{\isacharprime}\ {\isasymRightarrow}\ occurs\ r\isactrlisub {\isadigit{1}}\ r{\isacharprime}\isanewline
\isaindent{do\ {\isacharparenleft}}{\isacharbar}\ Const\ n\ {\isasymRightarrow}\ return\ False\isanewline
\isaindent{do\ {\isacharparenleft}}{\isacharbar}\ App\ r\isactrlisub {\isadigit{3}}\ r\isactrlisub {\isadigit{4}}\ {\isasymRightarrow}\ do\ b\ {\isasymleftarrow}\ occurs\ r\isactrlisub {\isadigit{1}}\ r\isactrlisub {\isadigit{3}}{\isacharsemicolon}\isanewline
\isaindent{do\ {\isacharparenleft}{\isacharbar}\ App\ r\isactrlisub {\isadigit{3}}\ r\isactrlisub {\isadigit{4}}\ {\isasymRightarrow}\ do\ }{\isacharparenleft}if\ b\ then\ return\ True\ else\ occurs\ r\isactrlisub {\isadigit{1}}\ r\isactrlisub {\isadigit{4}}{\isacharparenright}\isanewline
\isaindent{do\ {\isacharparenleft}{\isacharbar}\ App\ r\isactrlisub {\isadigit{3}}\ r\isactrlisub {\isadigit{4}}\ {\isasymRightarrow}\ }done{\isacharparenright}\isanewline
done}
\end{quote}
  This is a simple recursive traversal of \isa{r\isactrlisub {\isadigit{2}}},
  except that in the variable case, the traversal continues if the
  variable has already been instantiated.
  Since the term on the heap may contain cycles the function can
  diverge. Also note that the check \isa{r\isactrlisub {\isadigit{1}}\ {\isacharequal}\ r\isactrlisub {\isadigit{2}}} is a
  pointer equality test, not structural equality.

  To formulate the correctness property of such a function, it is
  convenient to re-state the property ``\isa{r\isactrlisub {\isadigit{1}}} occurs in \isa{r\isactrlisub {\isadigit{2}}}'' as an inductive relation 
  \isa{{\isachardoublequote}occurs{\isacharunderscore}in\ {\isacharcolon}{\isacharcolon}\ heap\ {\isasymRightarrow}\ \isasymalpha\ rtrm\ {\isasymRightarrow}\ \isasymalpha\ rtrm\ {\isasymRightarrow}\ bool{\isachardoublequote}}.
  Then, the crucial property connects \isa{occurs} and \isa{occurs{\isacharunderscore}in}:
  \begin{quote}
  \isa{{\isacharparenleft}h{\isacharcomma}\ h{\isacharprime}{\isacharcomma}\ b{\isacharparenright}\ {\isasymin}\ {\isasymlbrakk}\ occurs\ r\isactrlisub {\isadigit{1}}\ r\isactrlisub {\isadigit{2}}\ {\isasymrbrakk}\ {\isasymLongrightarrow}\ get{\isacharunderscore}ref\ r\isactrlisub {\isadigit{1}}\ h\ {\isacharequal}\ Var\ c\ None\ {\isasymLongrightarrow}\ occurs{\isacharunderscore}in\ h\ r\isactrlisub {\isadigit{1}}\ r\isactrlisub {\isadigit{2}}\ {\isacharequal}\ b}
  \end{quote}
  Note that there is no assumption that the pointer structures
  are acyclic, which is implicit in the assumption that
  the call to \isa{occurs} terminates and returns  \isa{b}.
  As there is no structural induction principle that can be used here,
  and we must use induction over the computation of the function---in
  other words, fixed-point induction.
  
  The induction rule produced by our prototype implementation is given
  below. With this rule, the inductive proof of the correctness
  property is straightforward. Note how the premises of the rule
  correspond to the cases in the function definition.
  \smallskip
  \begin{center}\mbox{} \blockrule{0.9\textwidth}%
{\isa{{\isasymAnd}r\isactrlisub {\isadigit{1}}\ h\ n\ {\isasymsigma}{\isachardot}\ Var\ n\ {\isasymsigma}\ {\isacharequal}\ get{\isacharunderscore}ref\ r\isactrlisub {\isadigit{1}}\ h\ {\isasymLongrightarrow}\ P\ r\isactrlisub {\isadigit{1}}\ r\isactrlisub {\isadigit{1}}\ h\ h\ True}\\
\isa{{\isasymAnd}r\isactrlisub {\isadigit{1}}\ r\isactrlisub {\isadigit{2}}\ h\ n{\isachardot}\ r\isactrlisub {\isadigit{1}}\ {\isasymnoteq}\ r\isactrlisub {\isadigit{2}}\ {\isasymLongrightarrow}\ Var\ n\ None\ {\isacharequal}\ get{\isacharunderscore}ref\ r\isactrlisub {\isadigit{2}}\ h\ {\isasymLongrightarrow}\ P\ r\isactrlisub {\isadigit{1}}\ r\isactrlisub {\isadigit{2}}\ h\ h\ False}\\
\isa{{\isasymAnd}r\isactrlisub {\isadigit{1}}\ r\isactrlisub {\isadigit{2}}\ h{\isacharprime}\ y\ h\ n\ r{\isacharprime}{\isachardot}\isanewline
\isaindent{\ \ \ }r\isactrlisub {\isadigit{1}}\ {\isasymnoteq}\ r\isactrlisub {\isadigit{2}}\ {\isasymLongrightarrow}\ P\ r\isactrlisub {\isadigit{1}}\ r{\isacharprime}\ h\ h{\isacharprime}\ y\ {\isasymLongrightarrow}\ Var\ n\ {\isacharparenleft}Some\ r{\isacharprime}{\isacharparenright}\ {\isacharequal}\ get{\isacharunderscore}ref\ r\isactrlisub {\isadigit{2}}\ h\ {\isasymLongrightarrow}\ P\ r\isactrlisub {\isadigit{1}}\ r\isactrlisub {\isadigit{2}}\ h\ h{\isacharprime}\ y}\\
\isa{{\isasymAnd}r\isactrlisub {\isadigit{1}}\ r\isactrlisub {\isadigit{2}}\ h\ n{\isachardot}\ Const\ n\ {\isacharequal}\ get{\isacharunderscore}ref\ r\isactrlisub {\isadigit{2}}\ h\ {\isasymLongrightarrow}\ P\ r\isactrlisub {\isadigit{1}}\ r\isactrlisub {\isadigit{2}}\ h\ h\ False}\\
\isa{{\isasymAnd}r\isactrlisub {\isadigit{1}}\ r\isactrlisub {\isadigit{2}}\ h{\isacharprime}\ h\ r\isactrlisub {\isadigit{3}}\ r\isactrlisub {\isadigit{4}}\ b{\isachardot}\ App\ r\isactrlisub {\isadigit{3}}\ r\isactrlisub {\isadigit{4}}\ {\isacharequal}\ get{\isacharunderscore}ref\ r\isactrlisub {\isadigit{2}}\ h\ {\isasymLongrightarrow}\ P\ r\isactrlisub {\isadigit{1}}\ r\isactrlisub {\isadigit{3}}\ h\ h{\isacharprime}\ b\ {\isasymLongrightarrow}\ b\ {\isasymLongrightarrow}\ P\ r\isactrlisub {\isadigit{1}}\ r\isactrlisub {\isadigit{2}}\ h\ h{\isacharprime}\ True}\\
\isa{{\isasymAnd}r\isactrlisub {\isadigit{1}}\ r\isactrlisub {\isadigit{2}}\ h{\isacharprime}{\isacharprime}\ y\ h\ r\isactrlisub {\isadigit{3}}\ r\isactrlisub {\isadigit{4}}\ h{\isacharprime}\ b{\isachardot}\isanewline
\isaindent{\ \ \ }App\ r\isactrlisub {\isadigit{3}}\ r\isactrlisub {\isadigit{4}}\ {\isacharequal}\ get{\isacharunderscore}ref\ r\isactrlisub {\isadigit{2}}\ h\ {\isasymLongrightarrow}\isanewline
\isaindent{\ \ \ }P\ r\isactrlisub {\isadigit{1}}\ r\isactrlisub {\isadigit{3}}\ h\ h{\isacharprime}\ b\ {\isasymLongrightarrow}\ {\isasymnot}\ b\ {\isasymLongrightarrow}\ P\ r\isactrlisub {\isadigit{1}}\ r\isactrlisub {\isadigit{4}}\ h{\isacharprime}\ h{\isacharprime}{\isacharprime}\ y\ {\isasymLongrightarrow}\ P\ r\isactrlisub {\isadigit{1}}\ r\isactrlisub {\isadigit{2}}\ h\ h{\isacharprime}{\isacharprime}\ y}}%
{\isa{{\isacharparenleft}y{\isacharcomma}\ r\isactrlisub {\isadigit{1}}{\isacharcomma}\ r\isactrlisub {\isadigit{2}}{\isacharparenright}\ {\isasymin}\ {\isasymlbrakk}\ occurs\ h\ h{\isacharprime}\ {\isasymrbrakk}}\ \isa{{\isasymLongrightarrow}}\ \isa{P\ h\ h{\isacharprime}\ y\ r\isactrlisub {\isadigit{1}}\ r\isactrlisub {\isadigit{2}}}}
\end{center}
  \smallskip
  While this rule looks intimidating at first, our (limited) practical
  experience suggests that there is no way around it.
  Before automation was available, the definition of the function and
  the proof of a similar induction rule took about 450 lines of very
  technical proof script, which is more than the correctness
  proof itself.
  The situation for the rest of the imperative unification algorithm is
  similar.
  With our new approach, these manual proofs are no longer needed.
\end{example}
\vspace*{-4pt}%
\end{isamarkuptext}%
\isamarkuptrue%
\isadelimtheory
\endisadelimtheory
\isatagtheory
\isacommand{end}\isamarkupfalse%
\endisatagtheory
{\isafoldtheory}%
\isadelimtheory
\endisadelimtheory
\end{isabellebody}%

%
\begin{isabellebody}%
\def\isabellecontext{Discussion}%
\isadelimtheory
\endisadelimtheory
\isatagtheory
\isacommand{theory}\isamarkupfalse%
\ Discussion\isanewline
\isakeyword{imports}\ Option{\isacharunderscore}Monad\isanewline
\isakeyword{begin}%
\endisatagtheory
{\isafoldtheory}%
\isadelimtheory
\endisadelimtheory
\isamarkupsection{Discussion%
}
\isamarkuptrue%
\label{discussion}
\begin{isamarkuptext}%
\paragraph{Implementation.}

The implementation of our technique is still in a prototype stage.  It
works well with the examples like the ones presented in this paper
but lacks several user-friendly features that would be needed for
productive use, e.g., support for mutually recursive
definitions and pattern matching. Moreover, the HOLCF dependencies
should be reduced to a minimum.

\paragraph{Higher-order recursion.}

Currently, our approach expects a fixed set of constructs in
monadic terms: \isa{{\isasymguillemotright}{\isacharequal}}, conditionals, recursive calls, and
  constant expressions not involving recursive calls (cf.\ the
continuity rules in Fig.~\ref{fig:cont}).
This basically limits the functions that can be defined to a
first-order fragment, since recursive calls must be applied and cannot
  be passed to functionals like \emph{map} or
\emph{fold}. To support higher-order recursion, one must handle more constructs, such as
a monadic map combinator \emph{mapM}.  Then, a suitable continuity
rule must be known to the tool, and (optionally) a rule to be used in
the induction rule refinement step.  The format of such rules is not
yet entirely clear, and it is part of future work to see if this
extension is possible and helpful in practice.

\paragraph{Option type vs.\ domain predicate.}

With the ability to define non-terminating recursive functions in the
option monad, we effectively have a new and alternative technique for
defining partial functions.  This raises the question about the
relative merits of the two approaches, and whether one should be
preferred over the other.
While this question can only be answered by comparing the approaches
  in concrete applications, a
few general things can be said:
\begin{itemize}
\item 
The main difference between the techniques is the format of
the function itself: while the function package produces a total
function and a predicate dom that describes
the arguments $x$ where \isa{f\ x} ``makes sense'', the option method
produces a function that returns an option type, where partiality is 
explicit in the result. As a price for this more accurate model, the
recursive equation must be written monadically, to deal with the
bottom values that may arise in recursive calls.

\item

The function package constrains the recursive equations with
the domain predicate, 
whereas the option method produces unconditional equations.
Thus, partial functions defined using the option method can be used
with Isabelle's code generator.

\item

The function package provides special support for tail-recursive
definitions. If the function is tail-recursive, the unconditional
specification can be derived as a theorem even for partial functions.

The same functionality could be provided using the domain-theoretic
approach: We choose the identity monad (i.e., no monad at all), and
take the flat pcpo where \isa{x\ {\isasymsqsubseteq}\ y\ {\isacharequal}\ {\isacharparenleft}x\ {\isacharequal}\ u\ {\isasymor}\ x\ {\isacharequal}\ y{\isacharparenright}} where \isa{u} is an arbitrary but fixed value. 
Since the bind operation (which is just application) is not
continuous, we cannot define arbitrary functions. We can however
define tail-recursive functions, since they can be written without
bind.
Thus, tail-recursion can be seen as a special case of the monadic
approach.
\end{itemize}

\paragraph{Other Applications.}

  Up to now, only option and state-exception monads are supported by
  our approach.  The question arises whether our treatment can be
  transferred to recursive definitions in other monads, e.g.,
  continuation or resumption monads.  This is the subject of future work.

  However, even in the current state, the approach can be of use in
  scenarios other than Imperative HOL: For example, Thiemann and Sternagel
  \cite{ThiemannS09} use an error monad to formalize a simple XML
  parser. Termination of the parser is irrelevant to the rest of their
  development, so it is currently assumed as an axiom, in order to
  obtain unconditional equations. Using our approach, these equations
  should come for free, but without axioms.

\paragraph{Transfinite iteration.}

There is a variant of the construction, where the notion of chain is
generalized to arbitrary totally ordered sets instead of just
countable ones. Then we can replace the countable iteration by a
transfinite one, taking the least upper bound for each limit ordinal
(this is easy to define inductively). Then, the continuity proof
obligation for each function can be replaced by the weaker
monotonicity. Otherwise the definition procedure remains the same, as
monotonicity proofs and continuity proofs are automated in the same
way. The advantage is that it is easier to set up new monads to work
with our method, since one must only prove monotonicity of bind. On
the other hand, it requires that we have least upper bounds exist for
non-countable chains as well. But in the application of interest this
is easily satisfied, so it may turn out that this variant of the
construction is preferrable. Here, more work is needed to fill in the
details.%
\end{isamarkuptext}%
\isamarkuptrue%
\isadelimtheory
\endisadelimtheory
\isatagtheory
\isacommand{end}\isamarkupfalse%
\endisatagtheory
{\isafoldtheory}%
\isadelimtheory
\endisadelimtheory
\end{isabellebody}%

\section{Conclusion}

Monadic functions present a challenge to the automated definition
mechanisms based on well-founded recursion.
We have shown that by using the fixed-point theorem for complete
partial orders, such definitions can be made with surprising ease and
result in equations that need no domain condition.
Automatically generated custom induction rules make the resulting
functions convenient to use, without having to refer to the iterative
construction used internally.

While it is perhaps not surprising that domain theory is up to this
task, using it for monadic definitions is particularly convenient,
since conditions like continuity and admissibility hold by
construction, and the overhead of handling undefinedness is
absorbed by the monad.

\paragraph{Acknowledgments}
I want to thank Brian Huffman for pointing me to the alternative
construction using transfinite iteration, and Lukas Bulwahn for
feedback on a draft of this paper and the implementation.

\bibliographystyle{eptcs}
\bibliography{bibliography}

\begin{thebibliography}{10}
\providecommand{\bibitemstart}[1]{\bibitem{#1}}
\providecommand{\bibitemend}{}
\providecommand{\bibliographystart}{}
\providecommand{\bibliographyend}{}
\providecommand{\url}[1]{\texttt{#1}}
\providecommand{\urlprefix}{Available at }
\providecommand{\bibinfo}[2]{#2}
\bibliographystart

\bibitemstart{nipkow_rewriting}
\bibinfo{author}{Franz Baader} \& \bibinfo{author}{Tobias Nipkow}
  (\bibinfo{year}{1998}): \emph{\bibinfo{title}{Term Rewriting and All That}}.
\newblock \bibinfo{publisher}{Cambridge University Press}.
\bibitemend

\bibitemstart{coqrecdef}
\bibinfo{author}{Gilles Barthe}, \bibinfo{author}{Julien Forest},
  \bibinfo{author}{David Pichardie} \& \bibinfo{author}{Vlad Rusu}
  (\bibinfo{year}{2006}): \emph{\bibinfo{title}{Defining and reasoning about
  recursive functions: a practical tool for the {Coq} proof assistant}}.
\newblock In: \bibinfo{editor}{Masami Hagiya} \& \bibinfo{editor}{Philip
  Wadler}, editors: {\sl \bibinfo{booktitle}{Functional and Logic Programming
  (FLOPS 2006)}}, {\sl \bibinfo{series}{Lecture Notes in Computer Science}}
  \bibinfo{volume}{3945}, \bibinfo{publisher}{Springer Verlag}, pp.
  \bibinfo{pages}{114 -- 129}.
\bibitemend

\bibitemstart{tphols2009}
\bibinfo{editor}{Stefan Berghofer}, \bibinfo{editor}{Tobias Nipkow},
  \bibinfo{editor}{Christian Urban} \& \bibinfo{editor}{Makarius Wenzel},
  editors (\bibinfo{year}{2009}): \emph{\bibinfo{title}{Theorem Proving in
  Higher Order Logics (TPHOLs 2009), 22nd International Conference, Munich,
  Germany, August 2009, Proceedings}}, {\sl \bibinfo{series}{Lecture Notes in
  Computer Science}} \bibinfo{volume}{5674}. \bibinfo{publisher}{Springer
  Verlag}.
\bibitemend

\bibitemstart{coqart}
\bibinfo{author}{Yves Bertot} \& \bibinfo{author}{Pierre Cast{\'e}ran}
  (\bibinfo{year}{2004}): \emph{\bibinfo{title}{Interactive Theorem Proving and
  Program Development: {Coq'Art}: The Calculus of Inductive Constructions}}.
\newblock \bibinfo{series}{Texts in theoretical computer science}.
  \bibinfo{publisher}{Springer Verlag}.
\bibitemend

\bibitemstart{Bertot2008}
\bibinfo{author}{Yves Bertot} \& \bibinfo{author}{Vladimir Komendantsky}
  (\bibinfo{year}{2008}): \emph{\bibinfo{title}{Fixed point semantics and
  partial recursion in {Coq}}}.
\newblock In: {\sl \bibinfo{booktitle}{Principles and practice of declarative
  programming (PPDP '08)}}, \bibinfo{publisher}{ACM}, \bibinfo{address}{New
  York, NY, USA}, pp. \bibinfo{pages}{89--96}.
\bibitemend

\bibitemstart{bovecapretta05}
\bibinfo{author}{Ana Bove} \& \bibinfo{author}{Venanzio Capretta}
  (\bibinfo{year}{2005}): \emph{\bibinfo{title}{Modelling general recursion in
  type theory.}}
\newblock {\sl \bibinfo{journal}{Mathematical Structures in Computer Science}}
  \bibinfo{volume}{15}(\bibinfo{number}{4}), pp. \bibinfo{pages}{671--708}.
\bibitemend

\bibitemstart{imperativeFP}
\bibinfo{author}{Lukas Bulwahn}, \bibinfo{author}{Alexander Krauss},
  \bibinfo{author}{Florian Haftmann}, \bibinfo{author}{Levent Erk{\"o}k} \&
  \bibinfo{author}{John Matthews} (\bibinfo{year}{2008}):
  \emph{\bibinfo{title}{Imperative functional programming in {Isabelle/HOL}}}.
\newblock In: \bibinfo{editor}{Otmane Ait~Mohamed}, \bibinfo{editor}{C{\'e}sar
  Mu{\~n}oz} \& \bibinfo{editor}{Sofi{\`e}ne Tahar}, editors: {\sl
  \bibinfo{booktitle}{Theorem Proving in Higher Order Logics (TPHOLs 2008)}},
  {\sl \bibinfo{series}{Lecture Notes in Computer Science}}
  \bibinfo{volume}{5170}, \bibinfo{publisher}{Springer Verlag}, pp.
  \bibinfo{pages}{134--149}.
\bibitemend

\bibitemstart{Capretta2005}
\bibinfo{author}{Venanzio Capretta} (\bibinfo{year}{2005}):
  \emph{\bibinfo{title}{General recursion via coinductive types}}.
\newblock {\sl \bibinfo{journal}{Logical Methods in Computer Science}}
  \bibinfo{volume}{1}(\bibinfo{number}{2}), pp. \bibinfo{pages}{1--18}.
\bibitemend

\bibitemstart{ConstableM85}
\bibinfo{author}{Robert~L. Constable} \& \bibinfo{author}{N.~P. Mendler}
  (\bibinfo{year}{1985}): \emph{\bibinfo{title}{Recursive definitions in type
  theory}}.
\newblock In: \bibinfo{editor}{Rohit Parikh}, editor: {\sl
  \bibinfo{booktitle}{Logic of Programs}}, {\sl \bibinfo{series}{Lecture Notes
  in Computer Science}} \bibinfo{volume}{193}, \bibinfo{publisher}{Springer
  Verlag}, pp. \bibinfo{pages}{61--78}.
\bibitemend

\bibitemstart{dubois98}
\bibinfo{author}{Catherine Dubois} \& \bibinfo{author}{V{\'e}ronique~Vigui{\'e}
  Donzeau-Gouge} (\bibinfo{year}{1998}): \emph{\bibinfo{title}{A step towards
  the mechanization of partial functions: domains as inductive predicates}}.
\newblock In: {\sl \bibinfo{booktitle}{CADE-15 Workshop on mechanization of
  partial functions}}.
\bibitemend

\bibitemstart{Gordon79LCF}
\bibinfo{author}{Michael J.~C. Gordon}, \bibinfo{author}{Robin Milner} \&
  \bibinfo{author}{Christopher~P. Wadsworth} (\bibinfo{year}{1979}):
  \emph{\bibinfo{title}{Edinburgh {LCF}: A Mechanised Logic of Computation}},
  {\sl \bibinfo{series}{Lecture Notes in Computer
  Science}}~\bibinfo{volume}{78}.
\newblock \bibinfo{publisher}{Springer Verlag}.
\bibitemend

\bibitemstart{Greve09_defminterm}
\bibinfo{author}{David Greve} (\bibinfo{year}{2009}):
  \emph{\bibinfo{title}{Assuming termination}}.
\newblock In: {\sl \bibinfo{booktitle}{ACL2 Workshop Proceedings}}.
\bibitemend

\bibitemstart{HaftmannN2010}
\bibinfo{author}{Florian Haftmann} \& \bibinfo{author}{Tobias Nipkow}
  (\bibinfo{year}{2010}): \emph{\bibinfo{title}{Code generation via
  higher-order rewrite systems}}.
\newblock In: \bibinfo{editor}{M.~Blume}, \bibinfo{editor}{N.~Kobayashi} \&
  \bibinfo{editor}{G.~Vidal}, editors: {\sl \bibinfo{booktitle}{Functional and
  Logic Programming (FLOPS 2010)}}, {\sl \bibinfo{series}{Lecture Notes in
  Computer Science}} \bibinfo{volume}{6009}, \bibinfo{publisher}{Springer
  Verlag}, pp. \bibinfo{pages}{103--117}.
\bibitemend

\bibitemstart{Huffman2008}
\bibinfo{author}{Brian Huffman} (\bibinfo{year}{2008}):
  \emph{\bibinfo{title}{Reasoning with powerdomains in {Isabelle/HOLCF}}}.
\newblock In: \bibinfo{editor}{Otmane Ait~Mohamed}, \bibinfo{editor}{C{\'e}sar
  Mu{\~n}oz} \& \bibinfo{editor}{Sofi{\`e}ne Tahar}, editors: {\sl
  \bibinfo{booktitle}{TPHOLs 2008: Emerging Trends Proceedings}}, pp.
  \bibinfo{pages}{45 --56}.
\newblock \bibinfo{note}{Department of Electrical and Computer Engineering,
  Concordia University}.
\bibitemend

\bibitemstart{Huffman2009}
\bibinfo{author}{Brian Huffman} (\bibinfo{year}{2009}): \emph{\bibinfo{title}{A
  purely definitional universal domain}}.
\newblock In \bibinfo{editor}{\bibinfo{editor}{Berghofer}} et~al.
  \cite{tphols2009}, pp. \bibinfo{pages}{260--275}.
\bibitemend

\bibitemstart{krauss_phd}
\bibinfo{author}{Alexander Krauss} (\bibinfo{year}{2009}):
  \emph{\bibinfo{title}{Automating Recursive Definitions and Termination Proofs
  in Higher-Order Logic}}.
\newblock \bibinfo{type}{Ph.D. thesis}, \bibinfo{school}{Institut f{\"u}r
  Informatik, Technische Universit{\"a}t M{\"u}nchen},
  \bibinfo{address}{Germany}.
\bibitemend

\bibitemstart{Krauss09_funs}
\bibinfo{author}{Alexander Krauss} (\bibinfo{year}{2010}):
  \emph{\bibinfo{title}{Partial and nested recursive function definitions in
  higher-order logic}}.
\newblock {\sl \bibinfo{journal}{Journal of Automated Reasoning}}
  \bibinfo{volume}{44}(\bibinfo{number}{4}), pp. \bibinfo{pages}{303--336}.
\bibitemend

\bibitemstart{Megacz2007}
\bibinfo{author}{Adam Megacz} (\bibinfo{year}{2007}): \emph{\bibinfo{title}{A
  coinductive monad for prop-bounded recursion}}.
\newblock In: \bibinfo{editor}{Aaron Stump} \& \bibinfo{editor}{Hongwei Xi},
  editors: {\sl \bibinfo{booktitle}{Proceedings of the {ACM} Workshop
  Programming Languages meets Program Verification, {PLPV} 2007}},
  \bibinfo{publisher}{ACM}, \bibinfo{address}{New York, NY, USA}, pp.
  \bibinfo{pages}{11--20}.
\bibitemend

\bibitemstart{holcf}
\bibinfo{author}{Olaf M{\"u}ller}, \bibinfo{author}{Tobias Nipkow},
  \bibinfo{author}{David von Oheimb} \& \bibinfo{author}{Oscar Slotosch}
  (\bibinfo{year}{1999}): \emph{\bibinfo{title}{{HOLCF}={HOL}+{LCF}.}}
\newblock {\sl \bibinfo{journal}{Journal of Functional Programming}}
  \bibinfo{volume}{9}(\bibinfo{number}{2}), pp. \bibinfo{pages}{191--223}.
\bibitemend

\bibitemstart{tutorial}
\bibinfo{author}{Tobias Nipkow}, \bibinfo{author}{Lawrence~C. Paulson} \&
  \bibinfo{author}{Markus Wenzel} (\bibinfo{year}{2002}):
  \emph{\bibinfo{title}{Isabelle/{HOL} --- A Proof Assistant for Higher-Order
  Logic}}, {\sl \bibinfo{series}{Lecture Notes in Computer Science}}
  \bibinfo{volume}{2283}.
\newblock \bibinfo{publisher}{Springer Verlag}.
\bibitemend

\bibitemstart{Regensburger_phd}
\bibinfo{author}{Franz Regensburger} (\bibinfo{year}{1994}):
  \emph{\bibinfo{title}{{HOLCF}: Eine konservative {Erweiterung} von {HOL} um
  {LCF}}}.
\newblock \bibinfo{type}{Ph.D. thesis}, \bibinfo{school}{Technische
  Universit\"at M\"unchen}.
\bibitemend

\bibitemstart{Regensburger95}
\bibinfo{author}{Franz Regensburger} (\bibinfo{year}{1995}):
  \emph{\bibinfo{title}{{HOLCF:} Higher order logic of computable functions}}.
\newblock In: \bibinfo{editor}{E.~Thomas Schubert}, \bibinfo{editor}{Phillip~J.
  Windley} \& \bibinfo{editor}{Jim Alves-Foss}, editors: {\sl
  \bibinfo{booktitle}{Higher Order Logic theorem proving and its applications
  (TPHOLs '95)}}, {\sl \bibinfo{series}{Lecture Notes in Computer Science}}
  \bibinfo{volume}{971}, \bibinfo{publisher}{Springer Verlag}, pp.
  \bibinfo{pages}{293--307}.
\bibitemend

\bibitemstart{wfrec}
\bibinfo{author}{Konrad Slind} (\bibinfo{year}{1996}):
  \emph{\bibinfo{title}{Function definition in higher-order logic.}}
\newblock In: \bibinfo{editor}{Joakim von Wright}, \bibinfo{editor}{Jim Grundy}
  \& \bibinfo{editor}{John Harrison}, editors: {\sl \bibinfo{booktitle}{Theorem
  Proving in Higher Order Logics (TPHOLs '96)}}, {\sl \bibinfo{series}{Lecture
  Notes in Computer Science}} \bibinfo{volume}{1125},
  \bibinfo{publisher}{Springer Verlag}, pp. \bibinfo{pages}{381--397}.
\bibitemend

\bibitemstart{ThiemannS09}
\bibinfo{author}{Ren{\'e} Thiemann} \& \bibinfo{author}{Christian Sternagel}
  (\bibinfo{year}{2009}): \emph{\bibinfo{title}{Certification of termination
  proofs using {CeTA}}}.
\newblock In \bibinfo{editor}{\bibinfo{editor}{Berghofer}} et~al.
  \cite{tphols2009}, pp. \bibinfo{pages}{452--468}.
\bibitemend

\bibliographyend
\end{thebibliography}

\end{document}